\documentclass[twocolumn]{aastex62}

\renewcommand{\vec}[1]{\mathbf{#1}}
\newcommand{\pd}[2]{\frac{\partial{#1}}{\partial #2}}

\usepackage{amsmath}
\usepackage{wasysym}

\received{}
\revised{}
\accepted{}

\shorttitle{Stellar Winds shaping Planetary Magnetospheres}
\shortauthors{Das et al.}

\begin{document}

\title{MODELING STAR-PLANET INTERACTIONS IN FAR-OUT PLANETARY AND EXOPLANETARY SYSTEMS}

\author{Srijan Bharati Das}
\affil{Center of Excellence in Space Sciences India, Indian Institute of Science Education and Research Kolkata, Mohanpur 741246, India}
\affil{Department of Physical Sciences, Indian Institute of Science Education and Research Kolkata, Mohanpur 741246, India}
\affil{Department of Geosciences, Princeton University, NJ 08544, USA}

\author{Arnab Basak}
\affil{Center of Excellence in Space Sciences India, Indian Institute of Science Education and Research Kolkata, Mohanpur 741246, India}

\author{Dibyendu Nandy}
\affil{Center of Excellence in Space Sciences India, Indian Institute of Science Education and Research Kolkata, Mohanpur 741246, India}
\affil{Department of Physical Sciences, Indian Institute of Science Education and Research Kolkata, Mohanpur 741246, India}

\author{Bhargav Vaidya}
\affil{Center of Excellence in Space Sciences India, Indian Institute of Science Education and Research Kolkata, Mohanpur 741246, India}
\affil{Centre of Astronomy, Indian Institute of Technology Indore, Khandwa Road, Simrol, Indore 453552, India}

\correspondingauthor{Dibyendu Nandy}
\email{dnandi@iiserkol.ac.in}

\begin{abstract}

The outflowing magnetized wind from a host star shapes planetary and exoplanetary magnetospheres dictating the extent of its impact. We carry out three-dimensional (3D) compressible magnetohydrodynamic (MHD) simulations of the interactions between magnetized stellar winds and planetary magnetospheres corresponding to a far-out star-planet system, with and without planetary dipole obliquity. We identify the pathways that lead to the formation of a dynamical steady-state magnetosphere and find that magnetic reconnection plays a fundamental role in the process. The magnetic energy density is found to be greater on the night-side than that on the day-side and the magnetotail is comparatively more dynamic. Magnetotail reconnection events are seen to associated with stellar wind plasma injection into the inner magnetosphere. We further study magnetospheres with extreme tilt angles keeping in perspective the examples of Uranus and Neptune. High dipole obliquities may also manifest due to polarity excursions during planetary field reversals. We find that global magnetospheric reconnection sites change for large planetary dipole obliquity and more complex current sheet structures are generated. We discuss the implications of these findings for injection of interplanetary species and energetic particles into the inner magnetosphere, auroral activity and magnetospheric radio emission. This study is relevant for exploring star planet interactions in the solar and extra-solar systems.

\end{abstract}

\keywords{magnetohydrodynamics (MHD) -- magnetic reconnection -- stars: winds, outflows}

\section{Introduction}\label{sec:intro}
The fate of a planetary system rests in the hands of the harboring star. The principal mode of interaction of a star with the planets it hosts is via the stellar wind~\citep{2012ApJ...754L..26M,2016A&A...594A..95A}. The variability of the stellar wind strength, which is related to the stellar activity and position of the planet with respect to the star, is an important factor in dictating planetary effects. Adhering to the planetary classification convention (close-in or far-out) of~\citet{2013ApJ...766....9S} based on distance of the planet from the star, it has been found that planets in close-in orbits ($\apprle 0.1$ AU) with their host star are subject to electromagnetic as well as gravitational interactions, i.e. tidal effects~\citep{2000ApJ...533L.151C} which can make the planets move either inward or outward due to transfer of angular momentum~\citep{2014ApJ...795...86S,2017ApJ...847L..16S}. \citet{2009ApJ...704L..85C} showed that close-in star planet interactions (referred to as SPIs henceforth) prevent the expansion of the stellar coronal magnetic field and the acceleration of the stellar wind. In order to study SPIs for the case of close-in planets using numerical simulations, both the planet and its hosting star are needed to be included in the computational domain~\citep{2015ApJ...815..111S,2017ApJ...847L..16S,2009ApJ...704L..85C,2011ApJ...733...67C,2015A&A...578A...6M} as the magnetic fields of both the entities get modified as a result of the interaction. The plasmoid loops trapped in between the stellar and the planetary magnetic fields then tend to modify the coronal magnetic fields as well~\citep{2008ApJ...687.1339K,2009A&A...505..339L,2012ApJ...754..137M}. For the case of fairly distant planets ($\apprge 0.5$ AU, e.g. Earth-like and beyond), the coronal magnetic field of the star is hardly affected and only the planetary magnetosphere is expected to deform. For this scenario, the stellar wind can be approximated as an independent physical entity which affects the planetary dynamics~\citep{2004JGRA..10911210T,2015GeoRL..42.9103D}.

The planetary effects of SPIs are numerous [for a detailed review, see~\citet{2017haex.bookE..25S}]. The habitability of a planet depends on whether it is capable of retaining a sustainable atmosphere around itself~\citep{2001Sci...291.1939S,2007Sci...315..501B}. The planetary magnetosphere serves as an invisible cage protecting the atmosphere. If the stellar wind is strong enough to penetrate the magnetosphere via magnetic reconnections, it might erode away a significant part of the atmosphere from the planet~\citep{2017ApJ...843L..33G,2017ApJ...837L..26D,nortmann2019ground}. The dynamic ram pressure of the wind is crucial in determining the distance from the planet till which the reconnection events are possible. A larger ram pressure would permit a smaller magnetopause. Since the region outside the magnetopause is not shielded, any atmosphere (which we consider to be charge neutral) can be ionized and taken away by the energetic stellar wind. So, as the wind grows stronger, the ram pressure increases and therefore allows greater stellar wind penetration, firstly due to compression of the magnetosphere, and, secondly due to opening up more closed field lines. \citet{2010Icar..210..539Z} presented an analytical model for interaction between atmospheres of non-magnetized planets and stellar winds from main sequence M stars and was able to provide a prediction of the time-scale for complete atmospheric loss. The magnetized SPIs might lead to enhanced radio emissions in the deformed magnetosphere of the planet~\citep{2007P&SS...55..598Z}. If the interplanetary magnetic field is able to penetrate the planet as a consequence of the interaction, ohmic dissipation may also lead to planetary heating~\citep{2008ApJ...685..521L}.  

The reason for investigating interactions between Sun-like stars and Earth-like planets is based on practical applications. The activity of the Sun affects our planet and indirectly, our lives as well. The solar activity evolution is of utmost importance as it controls the features of the outgoing solar wind. How the Sun changes over time~\citep{2018MNRAS.476.2465O,2018ApJ...856...53P} is, therefore, relevant from the planetary perspective. Observations have shown that very strong modulations in the solar wind occur over a period of roughly $1.3$ years~\citep{1994GeoRL..21.1559R}. In the near vicinity of the planet, variations in the solar wind intensity also depend on the position of the planet in its own orbit~\citep{Gómez1993} as well as its orientation with respect to the Sun~\citep{2008GeoRL..3518103M}. Prominent fluctuations in the planetary geomagnetic field orientation occur within a time range of less than $10$ years to periods of about $10^6$ years~\citep{cox_doell_1964}. Although geomagnetic field reversal is a rare event, the time-scale of the duration of geomagnetic field reversals is only about $1000$ to $6000$ years~\citep{1999Natur.401..885G}. The polarity excursions vastly affect the nature of magnetospheric deformation and magnetic reconnections during the interaction with the solar wind. The consequent effects are relevant for space weather variations~\citep{2006LRSP....3....2S,2007LRSP....4....1P}, geomagnetic storms~\citep{2006JGRA..111.7S14D}, aurora formation~\citep{1998JGR...10317543L}, spacecraft missions~\citep{2014Natur.513..291K}, etc. 

Driven by the above motivations, we study the interactions between the wind from a Sun-like star and the magnetosphere of an Earth-like planet using three dimensional (3D) compressible magnetohydrodynamic (MHD) simulations. As discussed earlier, keeping in mind the separation distance ($\approx 1.0$ AU) between the Sun and the Earth, the system can be considered to be a ``far-out" system. Therefore, it is prudent to keep the star out of the computational domain and only consider interactions between an incoming stellar wind and the planetary magnetosphere which is embedded in an interplanetary medium. The wind is assumed to be a pure shock and may have its intrinsic magnetic field oriented either northward or southward with respect to the direction of the planetary dipole axis. We use different values of the shock speed which replicates the variability of the solar wind intensity over time. We also consider different inclinations [including the present Earth-like obliquity (referred to as ``tilt" henceforth)] of the planetary dipole axis with respect to the equatorial plane in order to mimic planets with highly tilted axes or stages of polarity excursions during geomagnetic field reversals. Our main aim here is to provide a parameter space study and analyze the nature of magnetic reconnections that take place for different configurations of the star-planet system.

The structure of the paper is as follows. In Section~\ref{sec:theo}, we enlist the compressible MHD equations which we use to simulate the plasma system along with estimates of the physical quantities that are used. The numerical setup and description of the entire computational domain are also briefly described in the same section. In Section~\ref{sec:res}, we present our results. In Section~\ref{sec:con}, we present the conclusions of the study.

\begin{table*}[ht]
\centering
\caption{List of physical parameters used in the reference simulation and their notations.}
\label{tab:params}
\begin{tabular}{|l|c|c|}
\hline
\textbf{Physical Quantity}  & \textbf{Notation} & \textbf{Value Used} \\ \hline
Density in ambient medium    & $\rho_{\text{amb}}$      & $1.5 \times 10^{-23}$ $\rm{g\thinspace cm^{-3}}$  \\ \hline
Pressure in ambient medium    & $p_{\text{amb}}$      & $1.35\times 10^{-3}$ $\rm{dyne\thinspace cm^{-2}}$  \\ \hline
Density in stellar wind     & $\rho_{\text{sw}}$       & $4 \rho_{\text{amb}}$  \\ \hline
Stellar wind velocity       & $v_{\text{sw}}$          & $11.8 \times 10^7$ $\rm{cm\thinspace s^{-1}}$     \\ \hline
Ambient medium velocity     & $v_{\text{amb}}$         & 0.0                              \\ \hline
Temperature                 & $T_0$             & $4 \times 10^4$  $\rm{K}$               \\ \hline
Adiabatic index                       & $\gamma$          & 5/3                             \\ \hline
Orbital separation     & $d_{\text{orb}}$             & 1.0 $\rm{AU}$                           \\ \hline
Planetary mass     & $M_{\text{pl}}$             & $5.972\times 10^{27}$ $\rm{g}$                           \\ \hline
Planetary radius     & $r_{\text{pl}}$             & $6.371\times 10^{8}$ $\rm{cm}$                           \\ \hline
Planetary dipole moment     & $B_0$             & 0.31 $\rm{G}$                           \\ \hline
Magnetospheric tilt angle    & $\Theta_{\rm pl}$             & $0^{\circ}$, $11^{\circ}$, $45^{\circ}$, $90^{\circ}$                           \\ \hline
Stellar wind magnetic field & $B_{\text{sw}}$          & $4 \times 10^{-5}$  $\rm{G}$             \\ \hline
Magnetic diffusivity & $\eta$          & $10^{13}$  $\rm{cm^2\thinspace s^{-1}}$            \\ \hline
\end{tabular}
\end{table*}

\section{Star-Planet Interaction Model}\label{sec:theo}\label{MHD}
The plasma system that we are simulating is governed by the adiabatic equation of state and the following set of resistive MHD equations:
\begin{eqnarray}
&&\pd{\rho}{t} + \nabla \cdot (\rho \vec{v}) = 0 \label{eq:mhd1} \\
&&\pd{\vec{v}}{t} + (\vec{v} \cdot \nabla)\vec{v} + \frac{1}{4 \pi \rho} \vec{B} \times (\nabla \times \vec{B}) + \frac{1}{\rho} \nabla P = \vec{g} \label{eq:mhd2} \\
&&\pd{E}{t} + \nabla \cdot [(E+P)\vec{v} - \vec{B}(\vec{v} \cdot \vec{B})+ (\eta \cdot \vec{J}) \times \vec{B}] = \rho \vec{v} \cdot \vec{g} \nonumber \label{eq:mhd3} \\
 \\
&&\pd{\vec{B}}{t} + \nabla \times (\vec{B} \times \vec{v}) + \nabla \times(\eta \cdot \vec{J}) = 0 \label{eq:mhd4}
\end{eqnarray}
The variables $\rho$, $\mathbf{v}$, $\mathbf{B}$, $P$, $E$ denote the density, velocity, magnetic field, pressure and total energy density respectively. The pressure $P$ is the sum total of the thermal pressure ($p_t$) and the magnetic pressure ($p_b=\mathbf{B}^2/8 \pi$). $E$ stands for the sum total of the internal energy density, kinetic energy density and magnetic energy density. The vector $\mathbf{g}$ is the acceleration experienced by the fluid due to the gravitational field of the planet. $\mathbf{J}$ is the current density given by $\nabla \times \mathbf{B}$, neglecting displacement current. In our study, for simplicity, we consider a finite and isotropic magnetic diffusivity $\mathbf{\eta}$ which is constant in space and time. Table~\ref{tab:params} shows typical values of all the physical quantities used in our reference simulation and their corresponding notations. As shown in the table, we choose a value of magnetic diffusivity which falls within the range of realistic values as shown in earlier studies \citep{1986JGR....91.8057L,1996ESASP.389..459W,1999JGR...10417357R}. 

\begin{figure}[ht]
		\begin{center}
		\includegraphics[width=1.0\columnwidth]{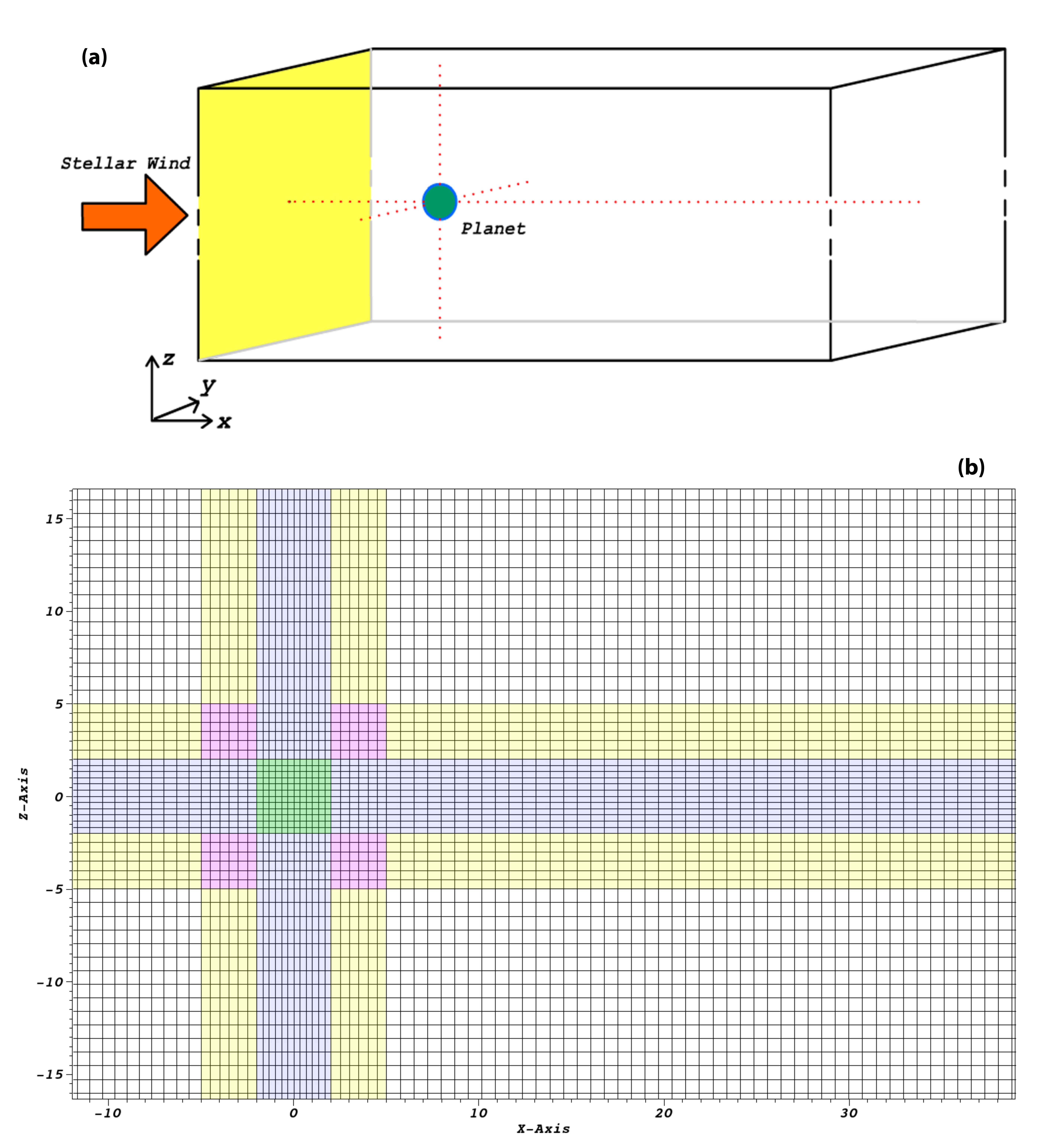}
		\caption{(a) Schematic diagram of the computational domain as used for star planet interaction simulation. The planet lies within the domain while the stellar wind comes in through the left boundary ($yz$-plane) with a velocity perpendicular to the plane of entrance. (b) Grid configuration in a zoomed-in portion of the computational box is shown ($xz$ slice at $y=0$). Static mesh refinement is implemented in all three directions near the vicinity of the planet of radius $1.0$ $R_{\rm E}$ placed at the origin (0,0,0). The grids marked by white, pink and green colors are square grids while those marked by yellow and blue colors are rectangular grids. Maximum grid resolution is obtained in a cube of dimension $4.0$ $R_{\rm E}$ containing the planet.}
		\label{fig:schematic}
		\end{center}
\end{figure}

In this study, we simulate the interaction of magnetized stellar wind with a planet hosting an intrinsic dipolar magnetic field. We use the 3D compressible MHD code \software{PLUTO \citep{2007ApJS..170..228M}} (ver. 4.3) to create a star-planet interaction module and solve the full set of resistive MHD equations [Eqs.~(\ref{eq:mhd1})-(\ref{eq:mhd4})] in a Cartesian box filled with magnetized ambient medium including a planet placed at the origin. For the Cartesian grid, the following axis convention was used: (a) $x$-axis is taken to be aligned with the line connecting the centres of the star and the planet with the positive $x$-direction directed along the line connecting the star to the planet, (b) $z$-axis is aligned with the zero-tilt dipole axis along the N to S magnetic poles and (c) $y$-axis is obtained by considering a right-handed Cartesian system with the information from $x$- and $z$-axes orientation. The origin $(0,0,0)$ is chosen to be the centre of the planet. Figure~\ref{fig:schematic}(a) shows a schematic representation of the computational domain and the coordinate convention chosen. The dashes at the midpoint of the edges of the box along the $z$-axis indicate that the dimensions are not to scale. The diagram simply serves the purpose of qualitatively representing the axial orientation for ease of understanding of the various directionalities that we use in the rest of the paper. We simulate our system until it reaches a steady state. We have used HLL Riemann solver with linear interpolation in space, MINMOD limiter and 2$^{\rm nd}$ order Runge-Kutta for temporal update. We impose the $\nabla \cdot \mathbf{B} = 0$ condition by using the divergence-cleaning method, an approach based on the generalized Lagrange multiplier (GLM) formulation~\citep{DEDNER2002645}. A constant value of  magnetic resistivity $\eta$ is used in our model as the causal mechanism for non-ideal processes like magnetic reconnections that are prevalent in regions of interaction of winds with the planetary atmosphere. The super-time-stepping scheme included in the code is implemented for this purpose~\citep{2017MNRAS.472.3147V}.

\subsection{Grid configuration}
The computational domain extends from $-45 R_{\rm E}$ to $225 R_{\rm E}$ in the $x$ direction, from $-25 R_{\rm E}$ to $25 R_{\rm E}$ in the $y$ direction and from $-225 R_{\rm E}$ to $225 R_{\rm E}$ in the $z$ direction. Static mesh refinement is implemented in all three directions in the 3D Cartesian grid configuration [please refer to Fig.~\ref{fig:schematic}(b)]. The planet of radius $1$ $R_{\rm E}$ has its centre at the origin (0,0,0). In any direction ($x$, $y$ or $z$), the region $[-2R_{\rm E},2R_{\rm E}]$ is resolved by $12$ grids, i.e. each grid resolves $0.33 R_{\rm E}$. Each of the regions $[-5R_{\rm E},-2R_{\rm E}]$ and $[2R_{\rm E},5R_{\rm E}]$ are resolved by $6$ grids each, i.e. each grid resolves $0.5 R_{\rm E}$. In the regions less than $-5 R_{\rm E}$ and greater than $5 R_{\rm E}$, each grid resolves about $0.73 R_{\rm E}$. The static mesh refinement gives rise to non-uniform rectangular grids in some parts of the computational domain as shown in Fig.~\ref{fig:schematic}(b). The grids in the vicinity of the planet are always square in shape.

\begin{figure*}[ht]
		\begin{center}
		\includegraphics[width=1.7\columnwidth]{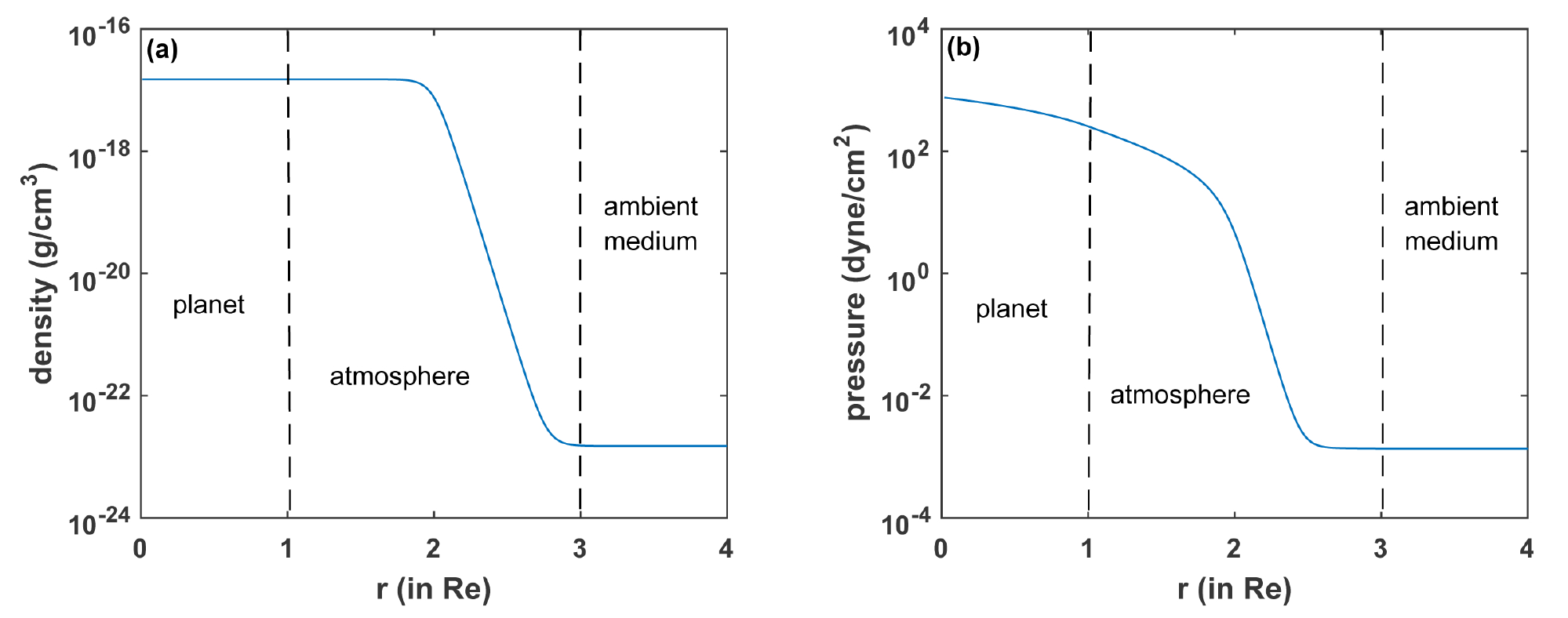}
		\caption{Density and pressure profiles considered as initial conditions for the simulations are shown in (a) and (b) respectively.}
		\label{fig:profiles}
		\end{center}
\end{figure*}

\begin{figure}[ht]
		\begin{center}
		\includegraphics[width=1.0\columnwidth]{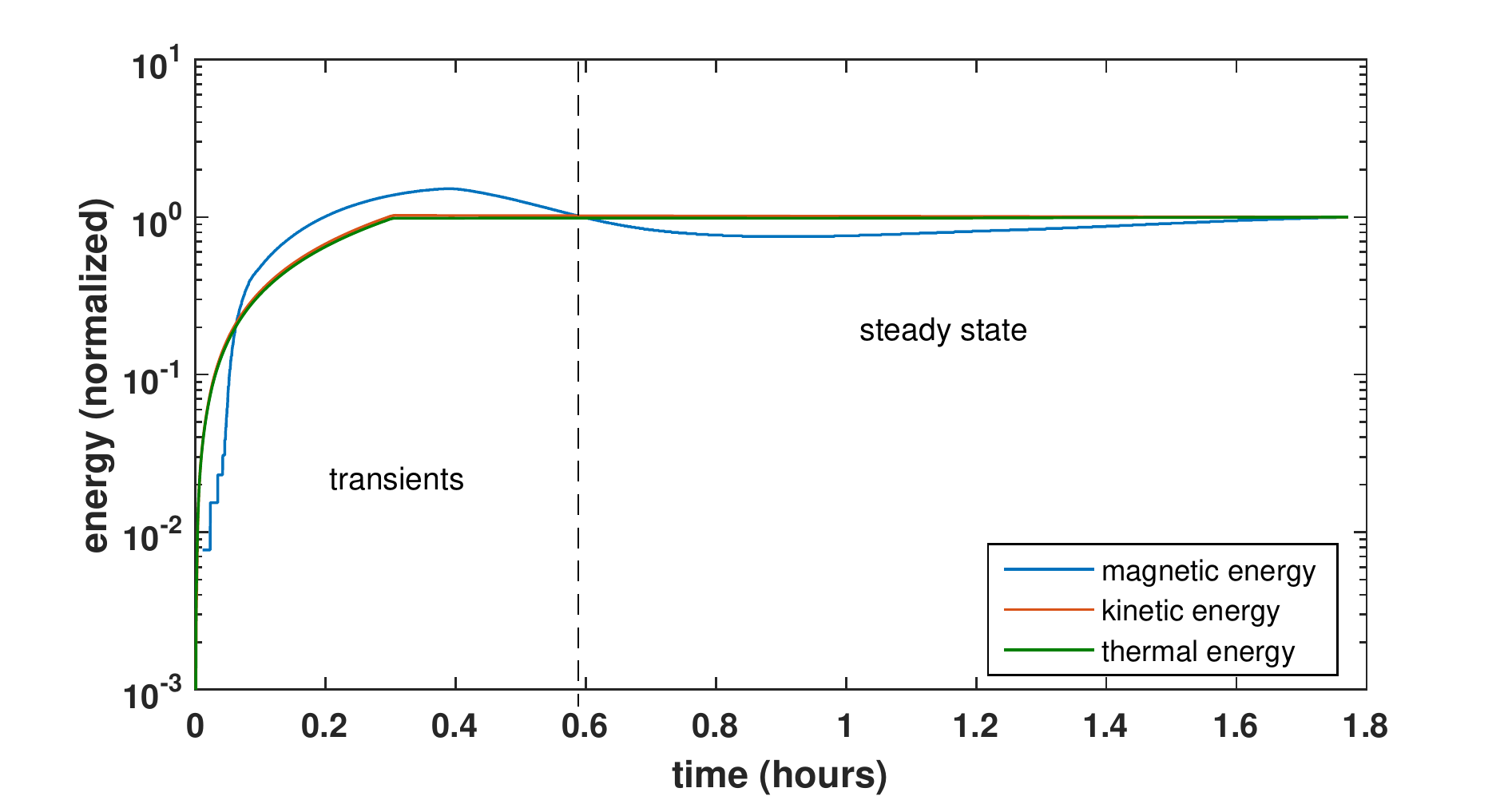}
		\caption{Attainment of steady state of the simulated star planet interaction module. The dynamical steady state is reached in about $30$ minutes of interaction as observed from the temporal evolutions of the global quantities e.g. kinetic, thermal and magnetic energies.}
		\label{fig:steady}
		\end{center}
\end{figure}

\subsection{The planet and it's atmosphere}
Within the sphere, defined by the planetary radius $r_{\rm pl}$, we keep a steady density profile ($\rho_{\rm pl}$) which is $10^6$ times the density of the ambient medium ($\rho_{\rm amb}$):


\begin{equation}
\rho_{\rm pl}=10^6 \rho_{\text{amb}}
\end{equation}
for $r \leq r_{\rm pl}$. Beyond the planetary radius, we set an atmosphere that extends upto $r=3r_{\rm pl}$. Density profile in the planetary atmosphere is chosen to ensure that a smooth transition exists between the density values at the boundaries $r=r_{\rm pl}$ and $r=3r_{\rm pl}$.
\begin{equation}
\rho_{\rm atm} (r) = \rho_{\rm pl} + \frac{(\rho_{\rm amb} - \rho_{\rm pl})}{2}\left[\tanh{\left\{9(r/r_{\rm pl}-2)\right\}}+1\right] \label{eq:atm_func}
\end{equation}
for $r_{\rm pl} \leq r \leq 3r_{\rm pl}$, where $\rho_{\rm pl}$ and $\rho_{\rm amb}$ are the densities of the planet and the ambient medium respectively. The above choice of density results in a planetary atmosphere with a mass equal to $2.8\times 10^{10}$ g.
The gravitational field in the entire computational domain is due to the actual mass of earth ($M_{\rm{pl}}=M_{\rm{E}}$) as given in Table~\ref{tab:params}. The mass is assumed to be a point mass kept at the centre of the planet (at $r=0$) and accordingly, the gravity is given by
\begin{equation}
g(r)= -\frac{GM_{\rm{pl}}}{r^2}
\end{equation}
The pressure distribution in the atmosphere is found out by numerical integration of the equation
\begin{equation}
\frac{dP}{dr} = -\rho_{\rm atm} (r) g (r) \label{eq:prs_diff}
\end{equation}
The pressure inside the planet is evaluated by extrapolating the value of pressure at the planet-atmosphere boundary as found from the numerical integration. Figure~\ref{fig:profiles} shows the density and pressure profiles that are fed as initial conditions. In the density profile in Fig.~\ref{fig:profiles}(a), the function~(\ref{eq:atm_func}) has been plotted in the atmosphere. The density in the planet ($\rho_{\rm pl}$), however, is not that of the actual Earth, but is kept continuous with the density of the atmosphere at the planet-atmosphere boundary so as to avoid any spurious simulation results due to sharp density jumps. The atmopsheric pressure profile in Fig.~\ref{fig:profiles}(b) is a result of numerical integration of equation~\ref{eq:prs_diff}, while that inside the planet is a simple extrapolation. It is important to note here that the planet is treated as an internal boundary and therefore, all the physical quantities inside the planet are essentially held at constant values as provided by the intial conditions. The above equations describe a gravitationally stratified atmosphere in hydrostatic balance with the planet and ambient medium. It is important to note here that the atmosphere so obtained is completely physical and represents that of an Earth-like planet.

A dipolar planetary magnetic field aligned along the $z$-axis is initialized throughout the domain . For zero-tilt ($\Theta_{\rm pl} = 0$), the magnetic dipole axis is aligned with the geographical axis that passes through the north and south pole of the planet. The following equations are used to initialize the three components of the dipolar magnetic field on the surface of the planet and outward:
\begin{eqnarray}
\label{dipole1} B_x &=& -B_0 r_{\rm pl}^3 \left(\frac{3 x z}{r^5}\right) \cos {\Theta_{\rm pl}} - B_0 r_{\rm pl}^3 \left(\frac{3 z^2 - r^2}{r^5}\right) \sin {\Theta_{\rm pl}} \nonumber \\
 \\
B_y &=& -B_0 r_{\rm pl}^3 \left(\frac{3 y z}{r^5}\right) \\
\label{dipole3} B_z &=& -B_0 r_{\rm pl}^3 \left(\frac{3 z^2 - r^2}{r^5}\right) \cos {\Theta_{\rm pl}} + B_0 r_{\rm pl}^3 \left(\frac{3 x z}{r^5}\right) \sin {\Theta_{\rm pl}} \nonumber \\
\end{eqnarray}
Here, $B_0$ denotes the magnetic dipole moment, $B_x, B_y, B_z$ denote the three components of the 3D magnetic field vector, $r$ is the radial distance from the center of the planet and $r_{\rm pl}$ is the radius of the earth-like planet (same size as that of Earth $R_{\text{E}}$ = 6.37$\times 10^8$ cm). The static mesh refinement is shifted by a small offset to avoid singularities in the magnetic fields at the origin $x=y=z=0$. The unperturbed planetary dipole magnetic field is applied as background field and additionally, the region inside the planet is treated as an internal boundary where the dynamical equations are not evolved. For the purpose of this study, we use  $\Theta_{\rm pl}$ = $0^{\circ}$, $11^{\circ}$, $45^{\circ}$, and $90^{\circ}$. It is to be noted here that for non-zero dipole tilt, the magnetic dipole axis of the planet does not pass through its geographic north and south poles but with an angle $\Theta_{\rm pl}$ relative to its geographic axis using left hand direction. Within the planet, the magnetic field is not allowed to evolve over time. This allows us to keep a steady planetary dipole throughout the duration of our simulation as we are not considering the planetary dynamo action in our simulations.

\subsection{Ambient medium}
We refer to the region $r > 3r_{\rm pl}$ to be the ambient medium. For modeling the initial magnetic field in this region, we use the same expression as the planetary dipole [see Eqs.~(\ref{dipole1}) through (\ref{dipole3})]. The density is initialized throughout the ambient medium to be $\rho_{\rm amb} = \rho_{\rm sw}$ while the pressure is set to the thermal pressure in this region as given in Table~\ref{tab:params} for $p_{\rm amb}$. We start our simulation with the planet, and its perfectly dipolar magnetic field in a static interplanetary medium. The evolution of the plasma properties of the ambient medium and consequent magnetospheric changes are induced due to the impact of the magnetized stellar wind. 

\subsection{Stellar wind}
The stellar wind is injected from the left boundary perpendicular to $yz$-plane ($x = -45 R_{\rm E}$). For all the other boundary faces of the Cartesian box, we implement the free flowing boundary conditions. The wind may have its magnetic field oriented either southwards (S-IMF) or northwards (N-IMF) with respect to the planetary geographical axis (along $z$ direction). The input parameters at the stellar wind injection boundary are obtained by solving the Rankine-Hugoniot jump conditions for the given shock velocity.

\begin{figure}[ht]
		\begin{center}
		\includegraphics[width=1.0\columnwidth]{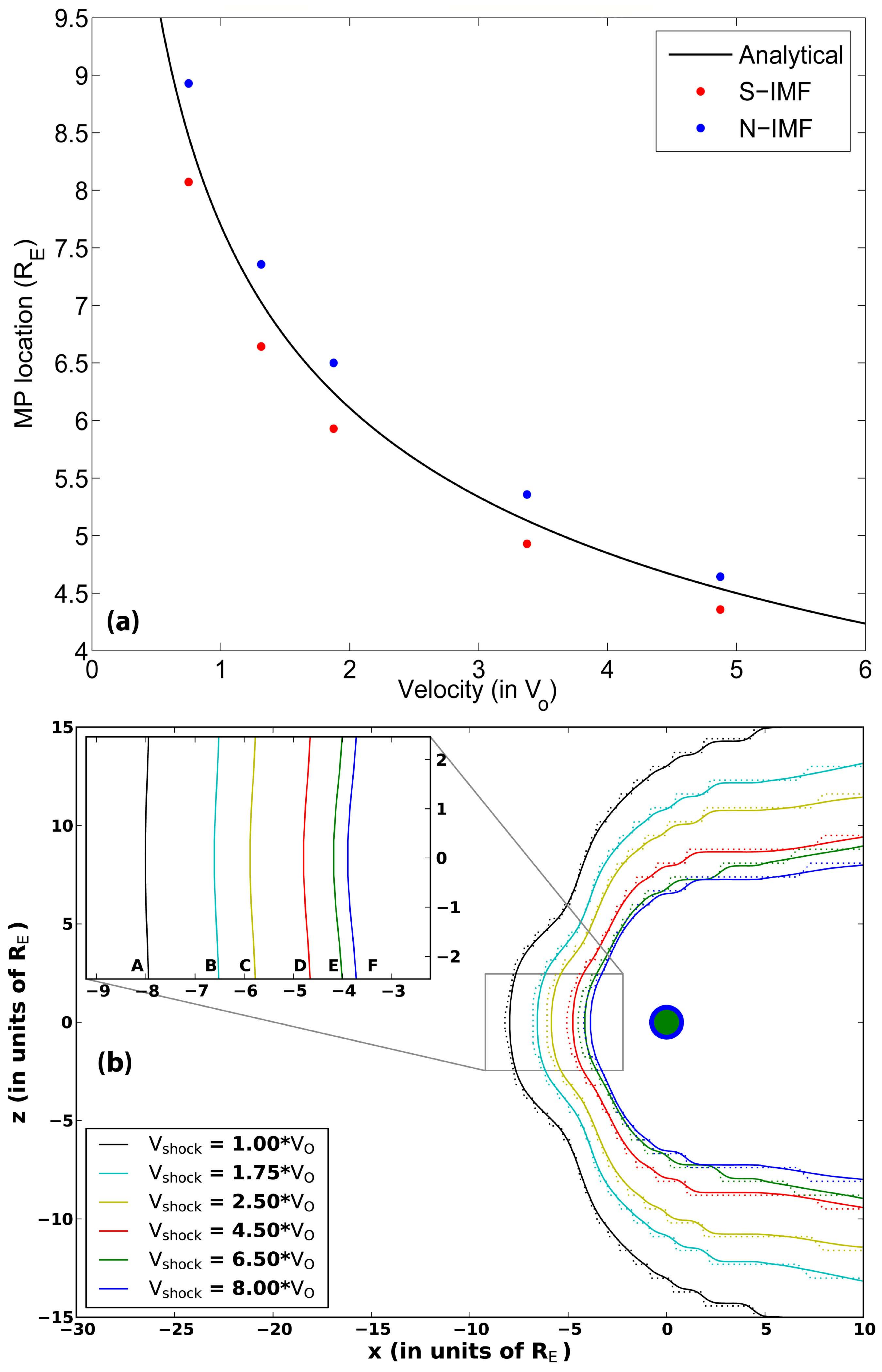}
		\caption{(a) Variation of magnetopause (MP) stand-off distance (from the planet) with the velocity of the stellar wind. The solid line denotes the expected value from the analytical expression while the red and blue data points are results obtained from simulations of no-tilt magnetosphere for S-IMF and N-IMF cases respectively. (b) Variation of magnetopause shape with velocity of the stellar wind. The cases A to F are for slowest to quickest winds respectively. $V_0$ = $350$ km/s is the slow wind speed at 1 AU.}
		\label{fig:magneto_vel}
		\end{center}
\end{figure}

\section{Results}\label{sec:res}
\subsection{Benchmarking the model: Magnetopause stand-off distance}
As a part of testing the robustness of the model, we carry out some sanity checks and reproduce certain expected features. Firstly, we define the steady state attained by the system in a particular simulation. Figure~\ref{fig:steady} shows the temporal evolutions of global quantities such as kinetic, thermal and magnetic energies normalized by their respective equilibrium values at $t=76$ minutes. It is observed that all the quantities attain steady state after about $30$ minutes of interaction. The magnetic energy plotted here is due to the deviation in magnetic field from the unperturbed planetary dipole profile. The primary feature of the magnetosphere that dictates the protective environment around the planet is the magnetopause. The magnetopause surface is demarcated by the zones where the pressure balance condition is satisfied between the ram pressure of the wind and the magnetic pressure of the planetary field. According to the model suggested in~\cite{1964JGR....69.1181M}, we use the following expression of sub-stellar magnetopause standoff distance~\citep{1998AnGeo..16..388P}:

\begin{equation} \label{mag_stand}
r_0 = \bigg(\frac{f^2 \mu_{\rm pl}^2}{8k\pi n m_{\rm p} v^2} \bigg)^{1/6}
\end{equation}

where, $r_0$ is the geocentric distance of the sub-stellar magnetopause stand-off distance, $f$ is used to account for the strengthening of planetary dipolar field due to the magnetopause currents, $k$ is a correction factor introduced by~\cite{1968ASSL...10..301S} to account for the nature of interaction of stellar wind with magnetopause and has been successively used in other studies \citep{1998AnGeo..16..388P}. The value of $k$ as stated in \cite{1968ASSL...10..301S} is 0.88. The parameter $\mu_{\rm pl}$ stands for the planetary dipole moment, $n$ and $v$ are the number density and uniform velocity of the stellar wind particles respectively and $m_{\rm p}$ is the mass of proton.

\begin{figure*}[ht]
		\begin{center}
		\includegraphics[width=1.8\columnwidth]{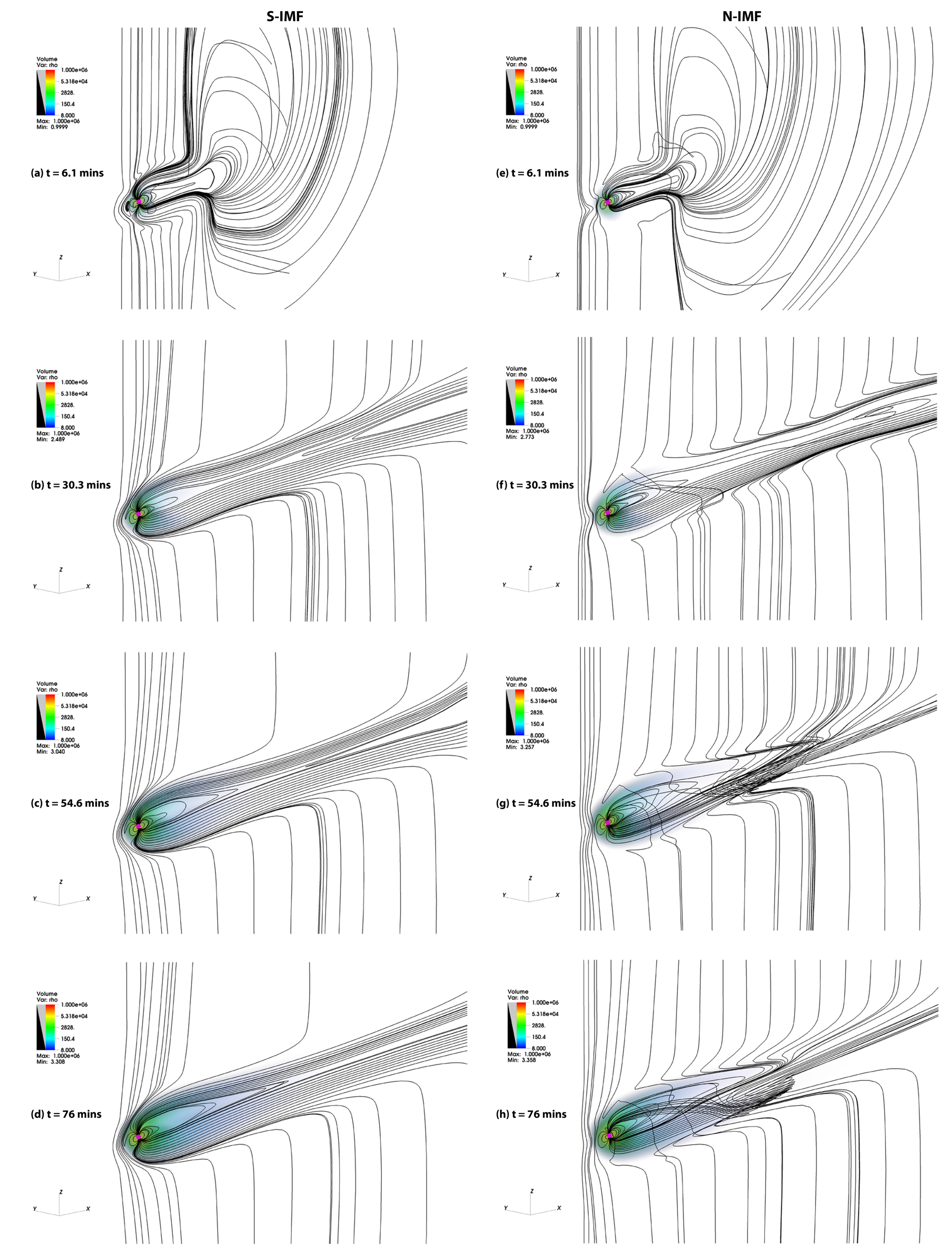}
		\caption{Temporal evolution of the planetary magnetosphere with $11^{\circ}$ inclination (Earth-like tilt) on the way to the steady state as a result of interaction with stellar wind for S-IMF and N-IMF cases. The colormap of the 3D volume rendering depicts density in units of $1.5\times 10^{-23}$ ${\rm g/cm^3}$.}
		\label{fig:evol}
		\end{center}
\end{figure*}

\begin{figure*}[ht]
		\begin{center}
		\includegraphics[width=1.8\columnwidth]{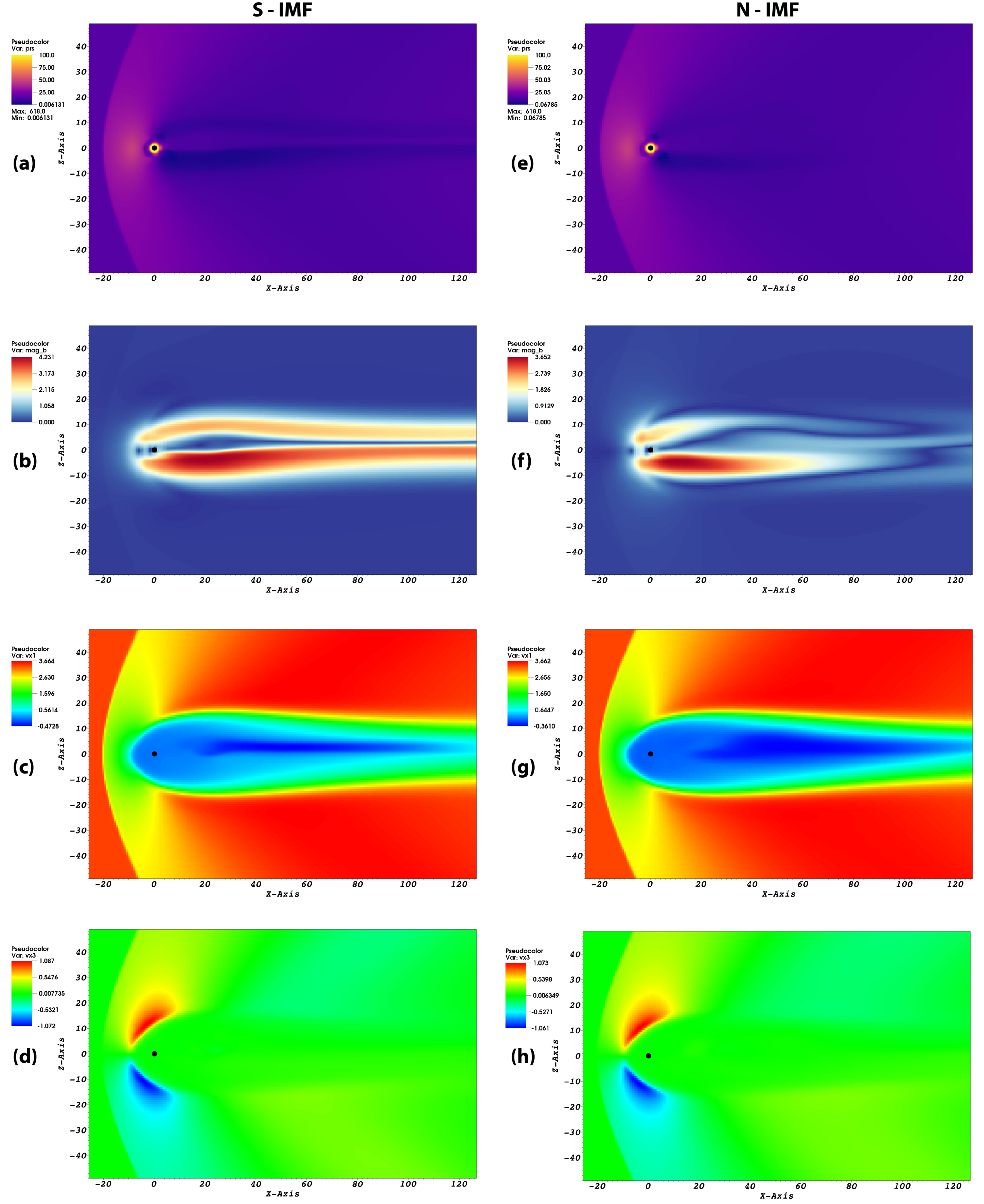}
		\caption{Steady state 2D profiles of pressure (top row, in units of $1.844\times 10^{-8}$ ${\rm dyne/cm^2}$), magnetic field strength (second row, in units of $4.814\times 10^{-4}$ G), and $x$ (third row) \& $z$ (fourth row) components of velocity (in units of $3.5\times 10^{7}$ cm/s) for S-IMF and N-IMF cases (Earth-like dipole tilt).}
		\label{fig:plots_2d}
		\end{center}
\end{figure*}

Figure~\ref{fig:magneto_vel}(a) shows a comparison between the expected values of magnetopause standoff distance obtained from its analytical expression [Eq.~(\ref{mag_stand})] and the results from our simulations. For this analysis, we have taken a range of stellar wind velocities obtained by varying shock speed between $V_{\rm shock} \approx V_0$ and $V_{\rm shock} \approx 8V_0$. Here, $V_0=350$ km/s is the value of slow solar wind at 1 AU. The S-IMF and N-IMF cases are plotted for the same wind speed. The plot shows that the results from both the N-IMF and S-IMF cases follow the same trend as shown by the analytical expression [Eq.~(\ref{mag_stand})]. However, for all the values of wind velocities, the stand-off distance obtained from simulating the S-IMF case is seen to be slightly lesser than the expected result. On the other hand, N-IMF data points are consistently slightly above the theoretically expected values~\citep{1998AnGeo..16..388P}. For S-IMF case, this can be attributed to the inward shift of magnetopause layer due to magnetic reconnection at the sub-stellar point which allows for greater penetration on the day-side magnetosphere. Whereas, for N-IMF, the outward migration is on account of the clustering of parallel field lines. Nevertheless, both the points for each wind velocity are very close to the expected value. The theoretically expected stand-off distance is approximately equal to the mean of the two simulated distances (from the N-IMF and S-IMF cases).

The trace of the magnetopause (in $xz$-plane) for different wind velocities for the S-IMF case is shown in Fig.~\ref{fig:magneto_vel}(b). As expected, we find that the magnetopause location migrates inwards towards the planet with an increase in wind velocity, which is also consistent with the result shown in Fig.~\ref{fig:magneto_vel}(a).

\begin{figure*}[ht]
		\begin{center}
		\includegraphics[width=1.9\columnwidth]{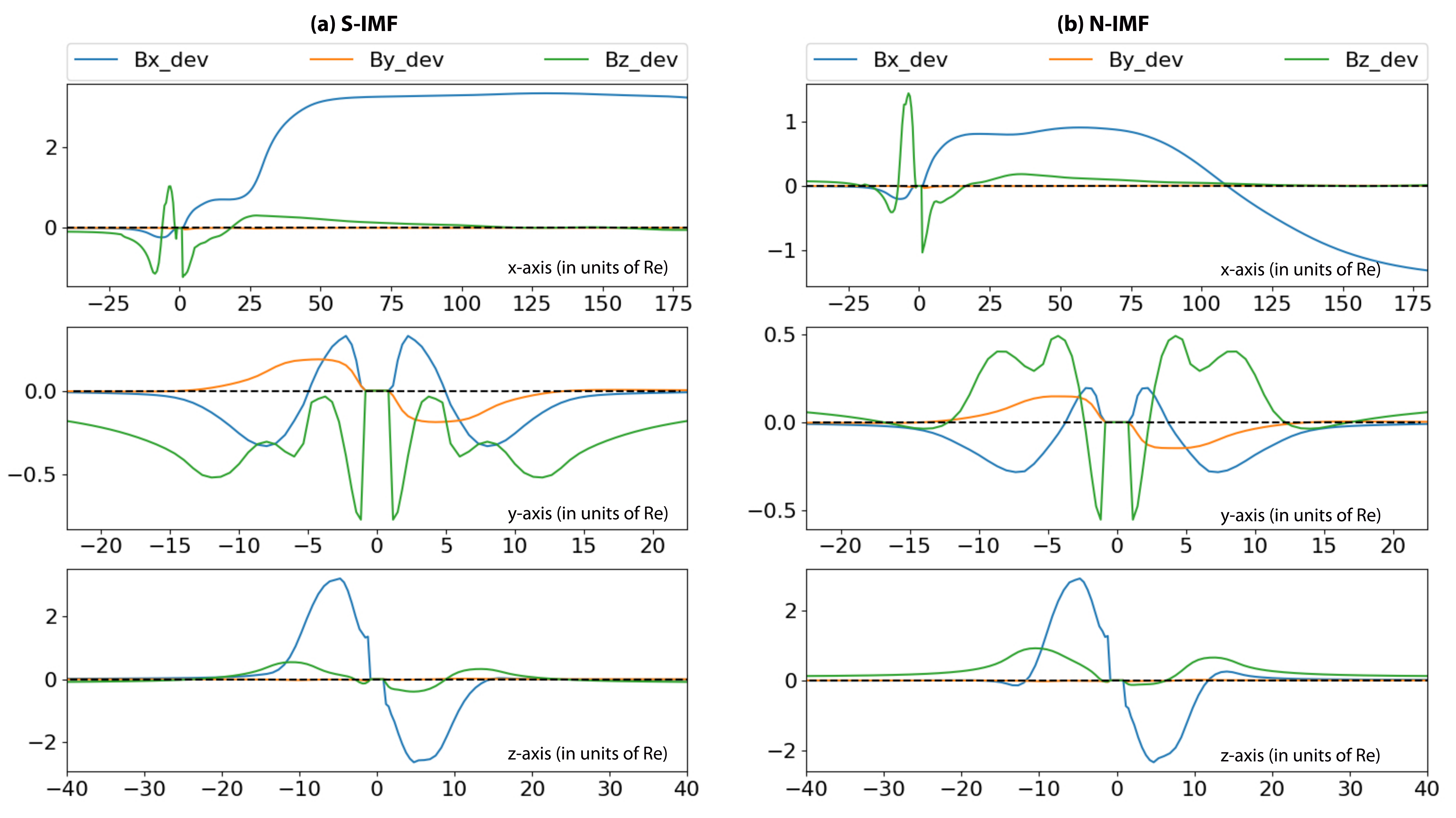}
		\caption{Variations of the three components of the magnetic field vector $\vec{B}\_{\rm dev}$ in $x$, $y$ and $z$ directions are shown for (a) S-IMF and (b) N-IMF cases respectively. $\vec{B}\_{\rm dev}$ represents the deviation in magnetic field from the unperturbed planetary dipole profile. The magnetic fields are in units of $4.814\times 10^{-4}$ G.}
		\label{fig:dev_solo}
		\end{center}
\end{figure*}

\begin{figure}[ht]
		\begin{center}
		\includegraphics[width=1.0\columnwidth]{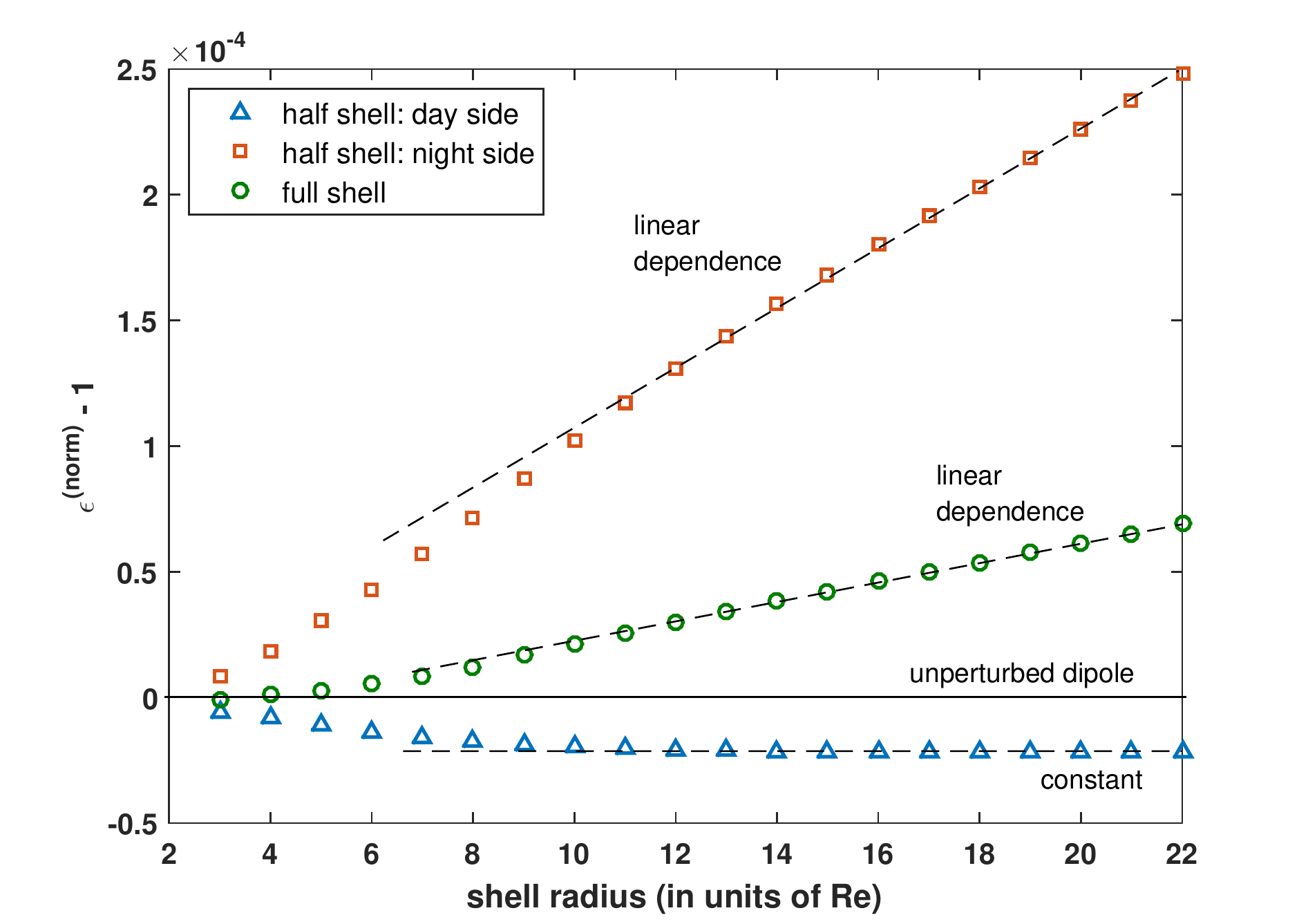}
		\caption{Normalized magnetic energy density as a function of radial distance from the centre of the planet for S-IMF case. The normalization is carried out with respect to the no-wind magnetic energy density.}
		\label{fig:mag_density}
		\end{center}
\end{figure}

\subsection{Magnetospheric dynamics for Earth-like tilt}
In this subsection, we present results of the effect of stellar wind on a planetary magnetosphere with Earth-like tilt for our reference simulation using $V_{\rm shock}=4.5V_0$ (corresponding parameters are given in Table~\ref{tab:params}). Furthermore, we study the possiblity of any plasma injection from the stellar wind into the planetary atmosphere which might be important for its evolution. We also investigate the extent of atmospheric mass loss as a result of the interaction.

\subsubsection{Evolution leading to establishment of steady state magnetosphere}
Here we present the temporal evolution of the interaction between the stellar wind and the planetary magnetosphere. We start from an unperturbed planetary dipole in the computational domain (as described in Section~\ref{sec:theo}) and impose the stellar wind at $t=0$. Figure~\ref{fig:evol} shows the magnetospheric evolution for S-IMF (left column) and N-IMF (right column) for the same time instants. The 3D volume rendering depicts the density while the magnetic field streamlines are plotted in black. It is to be noted here that while plotting the streamlines, their sources of origin have been kept on the $xz$-plane (please refer to Section~\ref{sec:theo} for coordinate convention used) with no constraint on their integration.

Let us discuss the S-IMF case first. At $t=6.1$ minutes [Fig.~\ref{fig:evol}(a)], the wind has just crossed the planetary magnetosphere. The opposite orientation of stellar and planetary field lines facilitates easy reconnections near the polar region. The night-side lobe in the far end remains unaffected. At $t=30.3$ minutes [Fig.~\ref{fig:evol}(b)], the wind has traversed the entire length of the computational box. The volume plot shows slow spread in density due to the interaction. On the night-side just behind the planet, a small magnetotail is formed due to pinching of the planetary field lines. This magnetic reconnection also results in the formation of plasma blobs that are advected out of the domain by the outgoing wind. The field lines leaving the upper and lower boundaries of the box originate from the planet due to reconnections. In Fig.~\ref{fig:evol}(c) ($t=54.6$ minutes), purely stellar field lines are now found at the right end. The system approaches steady state configuration in Fig.~\ref{fig:evol}(d) at $t=76$ minutes. The magnetotail stretches out longer in the night-side. Steady state 2D profiles of pressure, magnetic field strength and velocity for the S-IMF case are given in the left column of Fig.~\ref{fig:plots_2d}.

\begin{figure*}[ht]
		\begin{center}
		\includegraphics[width=2.0\columnwidth]{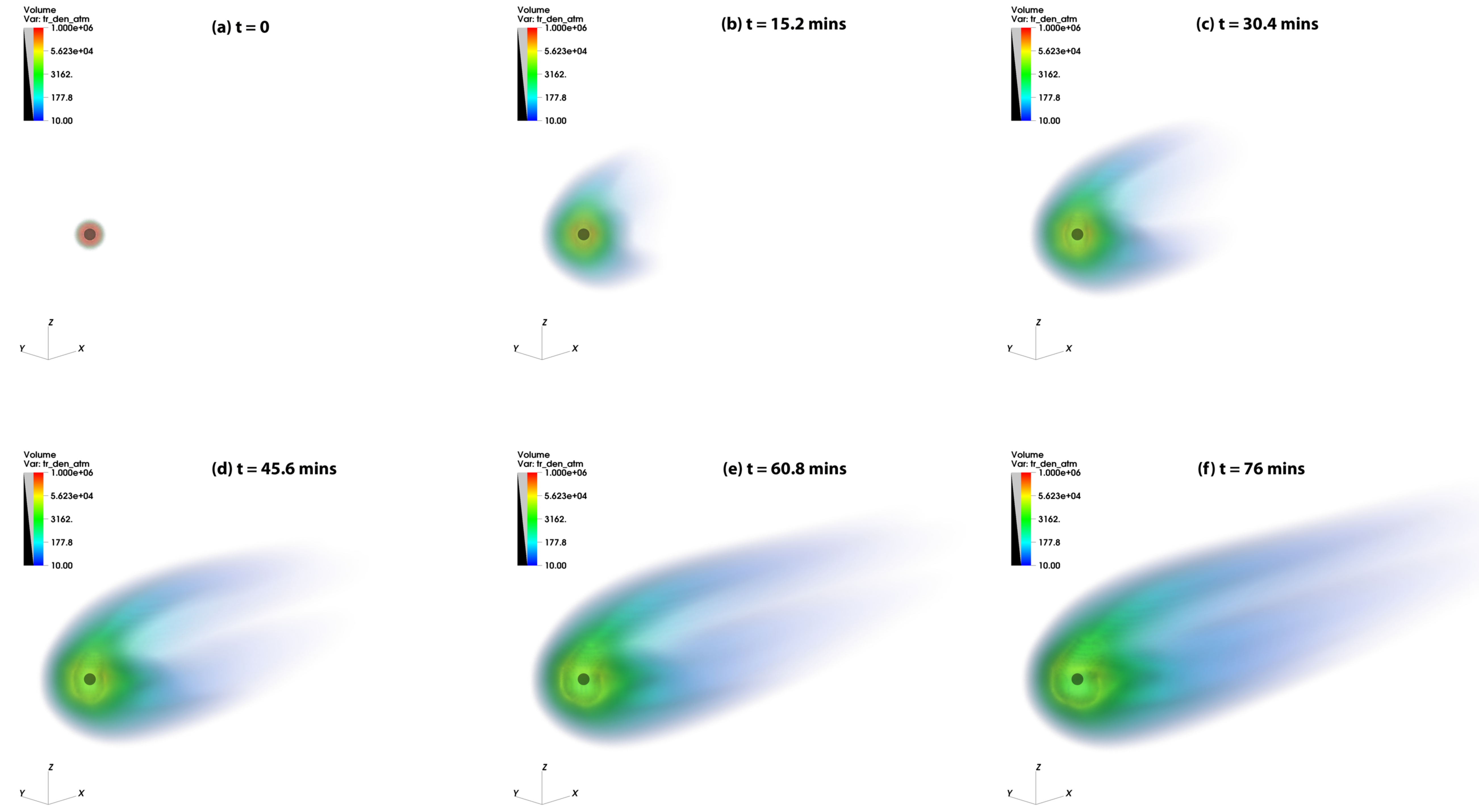}
		\caption{Temporal evolution of the planetary atmospheric mass loss as a result of interaction with stellar wind. The quantity plotted using 3D volume rendering is the density multiplied by the atmospheric passive scalar in units of $1.5\times 10^{-23}$ ${\rm g/cm^3}$.}
		\label{fig:loss_evol}
		\end{center}
\end{figure*}
\begin{figure*}[ht]
		\begin{center}
		\includegraphics[width=2.0\columnwidth]{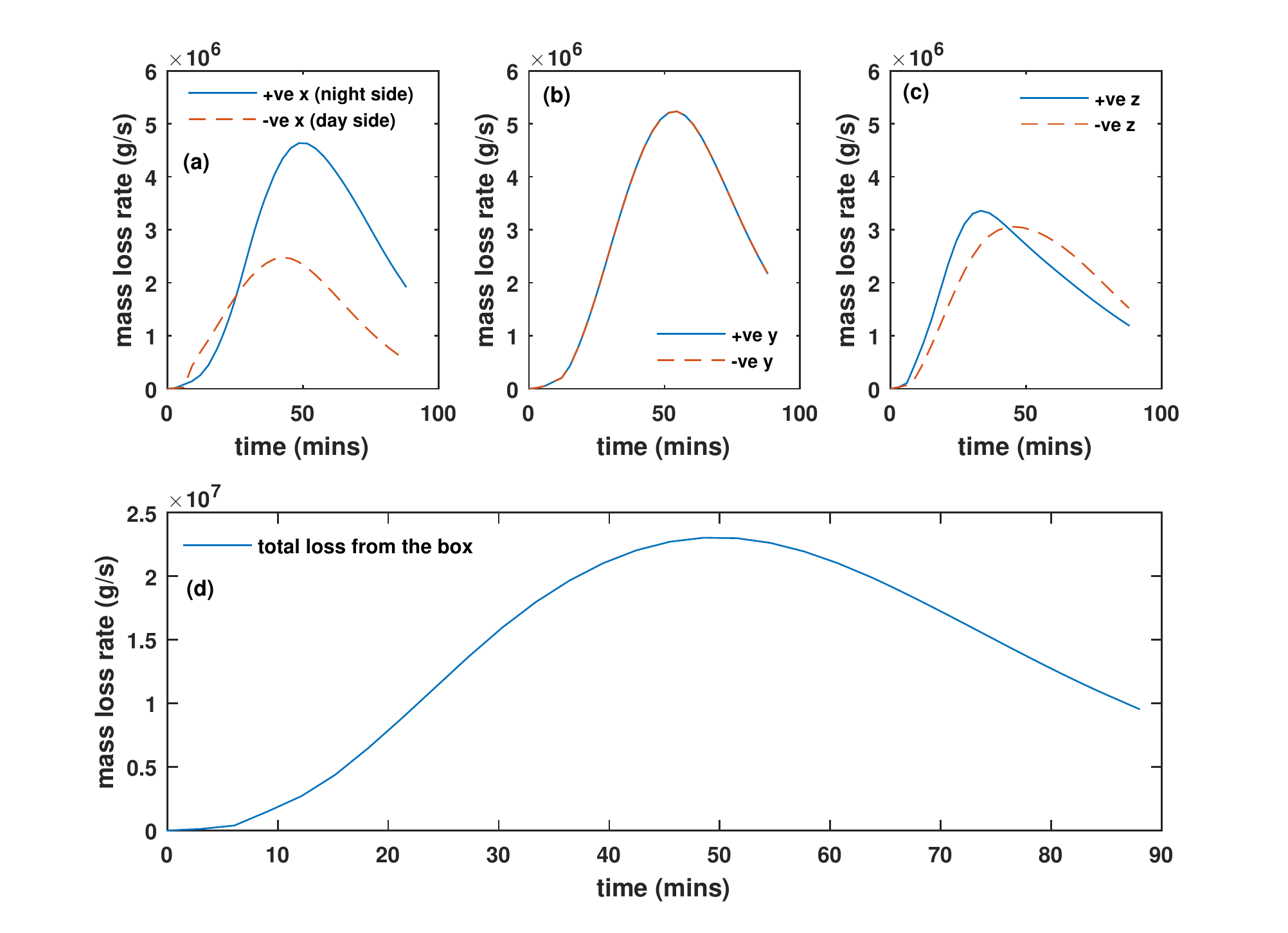}
		\caption{Atmospheric mass loss rates in the (a) x, (b) y, and (c) z directions and (d) total loss in all three directions. A cubic box with length extending from $-3.3R_{\rm E}$ to $+3.3R_{\rm E}$ in all three directions is considered for the mass loss rate calculation. The box encompasses the planet and its atmosphere.}
		\label{fig:loss_rates}
		\end{center}
\end{figure*}

The N-IMF case is more complicated. At $t=6.1$ minutes [Fig.~\ref{fig:evol}(e)], the day-side lobe gets compressed by the incoming wind and no reconnections occur in this region as the stellar and planetary field lines are oriented in the same direction. At $t=30.3$ minutes [Fig.~\ref{fig:evol}(f)], pinching of planetary field lines on the night-side gives rise to a magnetotail and a plasma blob at the far right end of the domain. The stellar field lines enter from the lower boundary, curl around the magnetosphere and leave the upper boundary. The reconnections of these stellar field lines occur with the upper and lower planetary lobes. At $t=54.6$ minutes [Fig.~\ref{fig:evol}(g)], the magnetosphere reaches a compact form being surrounded by purely stellar field lines. At the steady state in Fig.~\ref{fig:evol}(h) ($t=76$ minutes), the configuration remains almost similar to the earlier state. However, at the night-side near to the planet, the field lines get strongly twisted before leaving the domain. Steady state 2D profiles of pressure, magnetic field strength and velocity for the N-IMF case are given in the right column of Fig.~\ref{fig:plots_2d}. For both the S-IMF and N-IMF cases as presented in Fig.~\ref{fig:plots_2d}, we observe a bow-shock in front of the day-side magnetospheric lobe of the planet at about $-20R_{\rm E}$. The $x$ component of velocity decreases suddenly just right of the bow-shock and then increases gradually as we move further away from the planet in the night-side. The asymmetry about the equatorial plane in all the profiles is introduced due to the dipole tilt. Further illustration regarding the steady state field dynamics can be found in Section~\ref{subsec:steady_all}.

Figure~\ref{fig:dev_solo} shows the spatial distribution of the three components of the deviation ($\vec{B}\_{\rm{dev}}$) from the unperturbed dipolar magnetic field vector ($\vec{B}_{\rm{dipole}}$) as a result of the stellar wind interaction. $\vec{B}_{\rm{total}}$ = $\vec{B}_{\rm{dipole}}$ + $\vec{B}\_{\rm{dev}}$. For the $\vec{B}\_{\rm{dev}}$ profiles along $x$-axis, we keep $y=z=0$. Similarly, for plotting along $y$-axis, $x=z=0$ and for plotting along $z$-axis, $x=y=0$. Therefore, if a satellite carrying a magnetometer moves along the $x$-axis, the deviations of the total recorded magnetic field from the tilted dipole is expected to approximately match the profiles shown in the topmost plots in Fig.~\ref{fig:dev_solo}(a) and (b). Similarly, while moving along the $y-$ and $z-$axes, the profiles should resemble the middle and bottom plots respectively in Figs.~\ref{fig:dev_solo}(a) and (b). The profiles along $x-$axis show a considerable asymmetry about the origin (those along $y-$ or $z-$ are either a mirror image about the plane perpendicular to the respective axis or a mirror image about the origin). Along the $x-$axis, the deviation of the $x$ component of magnetic field $B{\rm x\_dev}$ is quite large in the night-side as compared to the deviation in the other two components. However, the profile of $B{\rm x\_dev}$ is entirely different for the S-IMF and N-IMF cases. Starting from $x = 0$ and moving along $+\hat{x}$, $B{\rm x\_dev}$ remains positive for S-IMF due to the open field lines at the far end [please see Fig.~\ref{fig:evol}(d)]. On the other hand for N-IMF, $B{\rm x\_dev}$ flips sign at the position where planetary field lines wrap around and purely stellar field lines begin at the right end [please see Fig.~\ref{fig:evol}(h)]. The deviations in the other two components of magnetic field are large near the vicinity of the planet in all three directions.

We also calculate magnetic energy density $\epsilon_{\rm{Earth-like}}$ for Earth-like case (S-IMF) with respect to an unperturbed dipole for no-wind $\epsilon_{\rm{no-wind}}$ as a function of radial distance from the centre of the planet. To facilitate the understanding of our output, we use a normalized energy density $\epsilon^{(norm)}$ = $\epsilon_{\rm{Earth-like}}$/$\epsilon_{\rm{no-wind}}$. For our analysis, we consider concentric spheres ($S_i$) of incremental radius $r = (3 + i)R_{\rm{E}}$ around the Earth-like planet with $i \in(0,19)$. So the innermost sphere has a radius of $3R_{\rm{E}}$ while the outermost is of radius $22 R_{\rm{E}}$. For the analysis of Fig.~\ref{fig:mag_density}, let us suppose that $\phi$ denotes the azimuthal angle with $\phi=0$ being along $+\hat{x}$ and $\phi=\pi$ along $-\hat{x}$. Now, we find the space-averaged magnetic energy density $\epsilon_i$ in each shell formed between two consecutive spheres ($S_i$ and $S_{i+1}$). We consider three kinds of shells: (a) $Sh_{\rm{day}}$: half shell in the day-side ($\pi<\phi<2\pi$), (b) $Sh_{\rm{night}}$: half shell in the night-side ($0<\phi<\pi$) and (c) $Sh_{\rm{total}}$: full shell ($0<\phi<2\pi$). This allows us to find the changes in relative contributions in energy density from the day-side sphere and night-side sphere as a function of radial distance. For $Sh_{\rm{day}}$, $\epsilon^{(norm)}-1$ decreases until around $12R_{\rm E}$ and then becomes almost constant in the region outside the magnetopause. For $Sh_{\rm{night}}$, magnetic energy density steadily increases and becomes proportional to radial distance [$(\epsilon^{(norm)}-1) \sim r$] from about $12R_{\rm E}$. Thus, contribution from day-side magnetic energy is much lower as compared to the night-side. So, the plot of $\epsilon^{(norm)}-1$ for a full shell, $Sh_{\rm{total}}$, follows linear dependence on shell radius just as $Sh_{\rm{night}}$ but is shifted to lower values due to the subdued contribution from day-side. Kindly note that for evaluating the contribution for the full shell, the sum of day- and night-side contributions have been normalized. On the day-side, the magnetospheric lobe is heavily compressed very near to the planetary surface due to the impact of the stellar wind. As a result, if we increase the shell radius, the magnetic energy density decreases and then eventually becomes constant outside the magnetopause where only the stellar wind magnetic field is present. However, on the night-side, planetary field lines extend long beyond the magnetotail due to advection by the wind, giving rise to relatively more complex and dynamic reconnections at the far right end. Thus, the magnetic energy density increases with increasing shell radius on the night-side.  

\subsubsection{Atmospheric mass loss and inner magnetospheric plasma injection}
The effects of SPIs are numerous. Important consequence are planetary atmospheric mass loss~\citep{2008P&SS...56.1260P,nortmann2019ground} and injection of stellar wind particles into the magnetosphere~\citep{2003JASTP..65..233L,2004physics..11066A,2017JGRA..122.2010H}. In our simulations, we have explored the possibility of plasma injection by analyzing large scale flows but not particle motion. We intialize the stellar wind and atmopshere with independent passive scalars normalized to unity and then, study the evolution of the stellar wind as well as the atmosphere. The atmosphere shows mass loss due to the interaction while inflow of wind plasma material is possible through the magnetotail.

Figure~\ref{fig:loss_evol} shows different stages in the phenomenon of atmospheric mass loss due to the impact of the stellar wind. The quantity plotted using 3D volume rendering is the product of density and atmospheric passive scalar. The frame at $t=0$ [Fig.~\ref{fig:loss_evol}(a)] shows the planet and its unperturbed atmosphere as the initial condition. As the incoming stellar wind interacts with the planetary magnetosphere, the atmopsheric matter is slowly dragged off in the night-side as depicted in successive frames [Fig.~\ref{fig:loss_evol}(b)-(f)] with increasing time. The asymmetry in the loss locations is introduced due the dipole tilt angle of $11^{\circ}$. Figure~\ref{fig:loss_rates} shows the temporal evolution of mass loss rates as obstained from the simulation. A cube of dimension $6.6R_{\rm E}$ (extending from $-3.3R_{\rm E}$ to $+3.3R_{\rm E}$ in all three directions) with its centre coinciding with the centre of the planet is considered. The box encompasses the planet and its atmosphere. Mass loss rates are calculated at all six faces of the cube as given in Figs.~\ref{fig:loss_rates}(a), (b) and (c). The total loss is plotted in Fig.~\ref{fig:loss_rates}(d). It is found that the mass loss rates increase first, reaches a peak value and then decreases steadily. The loss on the night-side is found to be greater than at the day-side as expected [Fig.~\ref{fig:loss_rates}(a)]. Since there is no aymmetry in the magnetospheric configuration in the $y$ direction, the loss rates are equal at both the faces [Fig.~\ref{fig:loss_rates}(b)]. In the $z$ direction, the asymmetry is introduced by the dipole tilt [Fig.~\ref{fig:loss_rates}(c)]. The atmospheric mass loss rate for hot Jupiter (which has a much larger radius and therefore, a larger surface area holding the atmosphere) was found by \cite{2008P&SS...56.1260P} to be about $3.5\times 10^{10}$ g/s. In our case (Earth-like), the peak value of mass loss rate reaches to about $2.3\times 10^{7}$ g/s.

The intensity of the magnetospheric radio emission can also be estimated which is proportional to the radio power given by
\begin{equation}
P_r = \delta\frac{\dot{M}_* v_{\rm sw}^2 R_{\rm eff}^2}{4a^2}
\end{equation}
as used by \cite{2005MNRAS.356.1053S} and \cite{2018MNRAS.479.1194D}. Here, $\delta$ is the efficiency parameter, $\dot{M}_*$ is the stellar mass loss rate, $R_{\rm eff}$ is the effective radius of the planetary magnetosphere as seen by the stellar wind and $a$ is the orbital radius. Here $\delta \sim 7\times 10^{-6}$, $\dot{M}_*$ for a Sun-like star is $\dot{M}_{\odot}\approx 1.268\times 10^{12}$ g/s, $R_{\rm eff}$ is the magnetopause stand-off distance which is roughly $5R_{\rm E}$, $a$ is $1$ AU and $v_{\rm sw}$ from Table~\ref{tab:params}. The calculation gives a $P_r$ value as given below in a scaled equation form.
\begin{eqnarray}
P_r = 1.4\times 10^{15} \thinspace {\rm erg\thinspace s^{-1}} \left( \frac{\delta}{7\times 10^{-6}} \right) \left( \frac{\dot{M}_*}{\dot{M}_{\odot}}\right) \nonumber\\
\left( \frac{v_{\rm sw}}{11.8\times 10^7\thinspace {\rm cm\thinspace s^{-1}}} \right)^2 \left(\frac{R_{\rm eff}}{5R_{\rm E}}\right)^2 \left(\frac{a}{1\thinspace {\rm AU}}\right)^{-2}
\end{eqnarray}
The above estimate for the radio power is for our reference simulation in which a stellar wind velocity of about 3 times the value of the typical slow solar wind velocity at 1 AU is chosen. Also, the magnetopause stand-off distance is half the typical value for our Sun-Earth system. Therefore, considering that the host star in our reference simulation has the same mass loss rate as the Sun, the above radio power estimate will be about 2 times higher than that expected for the Sun-Earth system at 1 AU. For the case of a hot Jupiter, \cite{2018MNRAS.479.1194D} estimated the radio power to be $1.42\times 10^{19}$ erg/s.

\begin{figure}[ht]
		\begin{center}
		\includegraphics[width=1.0\columnwidth]{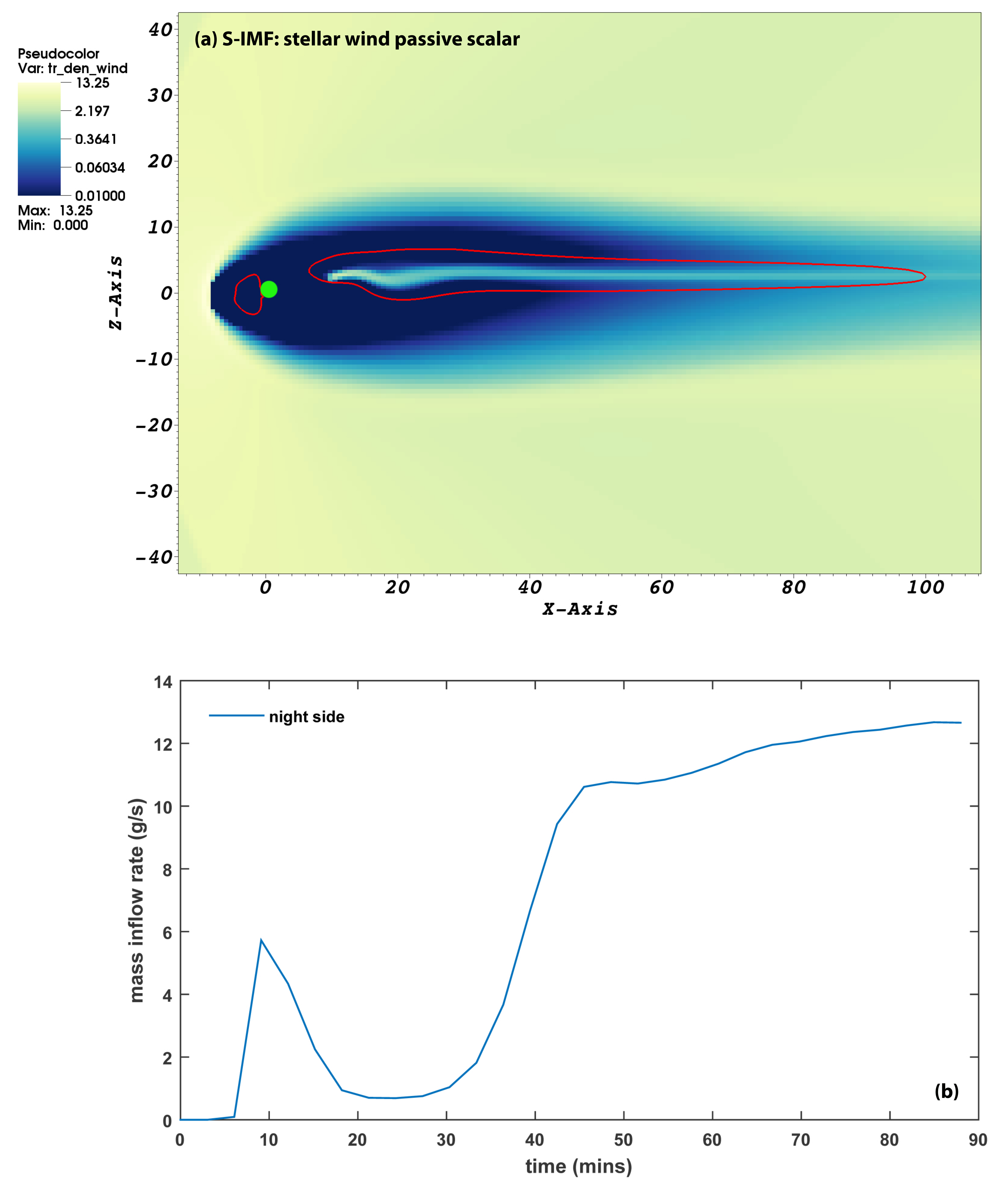}
		\caption{(a) Spatial variation of stellar wind passive scalar in the $xz$-plane after about $75$ minutes of interaction in the dynamical steady state regime for the S-IMF case. The red contour lines demarcate the regions of inflow towards the planet, i.e. negative values of the $x$ component of velocity. The inflow of stellar plasma material into the planetary atmosphere occurs through the magnetotail till about $10$ $R_{\rm E}$. (b) Temporal evolution of plasma mass inflow rate in the night-side at a distance of $10$ $R_{\rm E}$ from the centre of the planet.}
		\label{fig:inflow}
		\end{center}
\end{figure}

\begin{figure*}[ht]
		\begin{center}
		\includegraphics[width=2.0\columnwidth]{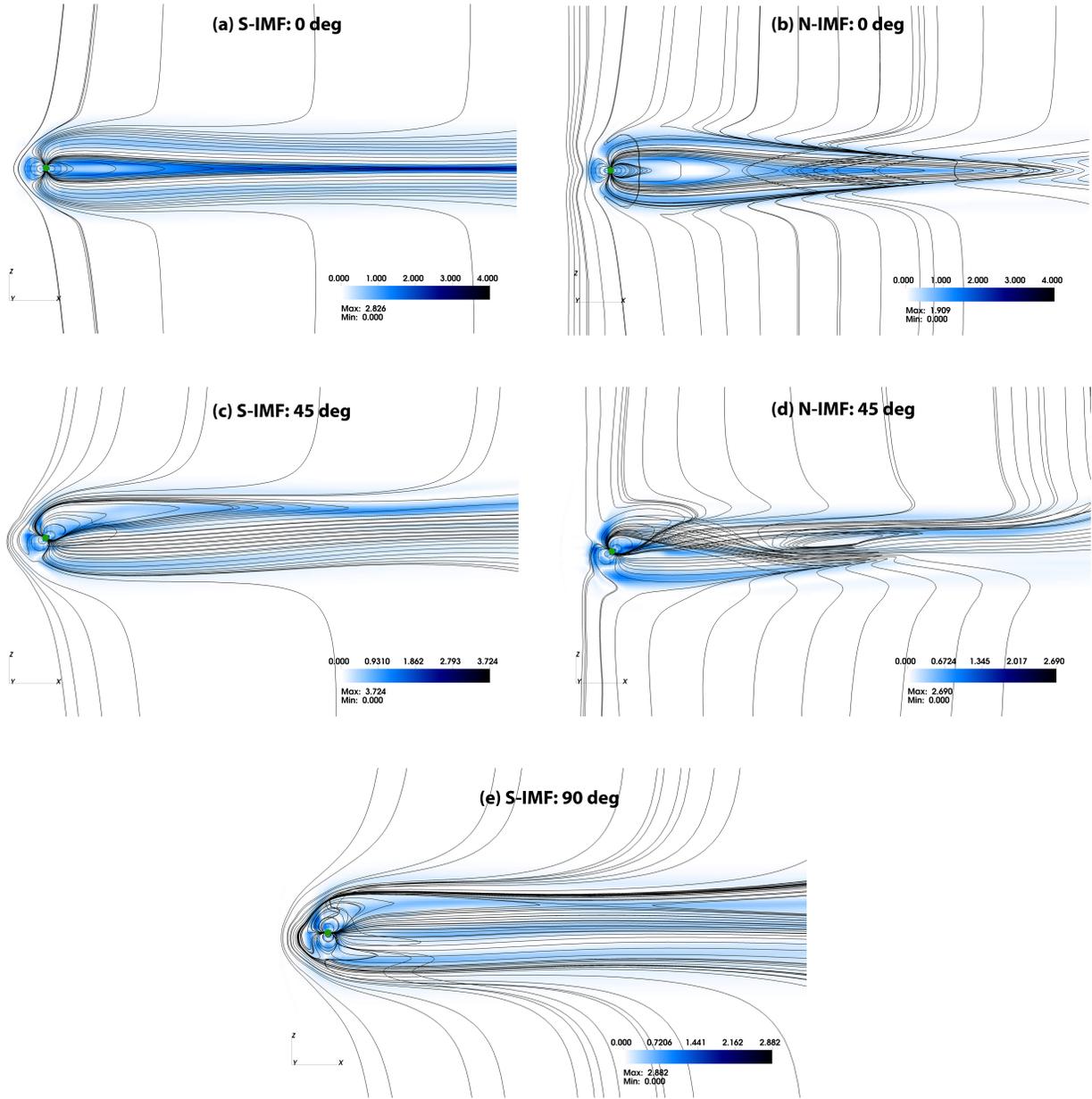}
		\caption{3D steady state magnetospheric configurations with no tilt and extreme tilt angles for both S-IMF and N-IMF cases. The N-IMF case for $90^{\circ}$ is not shown as it is just a flip about the horizontal of the configuration in S-IMF case. The background colormap shows the magnitude of the total current density ($\vec{j}$).}
		\label{fig:tilts_all_red}
		\end{center}
\end{figure*}

Figure~\ref{fig:inflow}(a) shows plot of the stellar wind passive scalar density in the $xz$-plane at $y=0$ for S-IMF case after about $76$ minutes of interaction. The red contour lines are used to denote $v_x=0$. So, the contours encompass the regions of negative velocity in the $x$-direction, i.e. inflow towards the planet through the magnetotail. This basically means that if any non-zero stellar wind passive scalar density portion lies within the contoured regions, there is a chance of stellar wind plasma getting injected into the planetary atmosphere as evident from the figure. This plasma can then be trapped in the closed field lines of the planet. The plasma particles can thereafter, exhibit the phenomenon of magnetic mirroring~\citep{kulsrud2005plasma} which results in its bouncing back and forth between the two poles of the planet. This continues so long as the particle trajectory satisfies an angle smaller than the pitch angle. In the instance where the angle between the particle's trajectory with the local magnetic field exceeds the pitch angle, these particles can escape into the surrounding media (which, near the poles, is the planetary atmosphere). Thus, particles reaching into near $10 R_{\rm E}$ or lesser, could potentially get trapped in the closed field lines, exhibit mirroring and a certain fraction could be lost to the planetary or exoplanetary atmosphere. It is to be noted here that the asymmetry induced due to the introduction of the tilt facilitates greater inflow through the magnetotail and polar regions. The curve in Fig.~\ref{fig:inflow}(b) shows the temporal evolution of stellar plasma influx rate through the magnetotail in units of g/s. For plotting the above, a square plane of dimension $10R_{\rm E}$ is placed at a distance of $10R_{\rm E}$ from the centre of the planet in the night-side and mass influx rate through that plane towards the planet due to the stellar wind passive scalar is calculated. It is found that after initial transients, the mass inflow rate reaches a steady value of about $12.5$ g/s.

\subsection{Magnetospheres with extreme tilt angles: Steady state reconnections\label{subsec:steady_all}}
Most planets in our solar system have a moderately tilted dipolar axis which is comparable to that of the Earth. Nevertheless, we also find planets where the dipole axis is extremely tilted, e.g Uranus and Neptune~\citep{2010SSRv..152..251R}. Such planets hosting an intrinsic dipolar magnetosphere have a convective core~\citep{1999Natur.401..885G} and the resulting geodynamo causes periodic reversals in the polarity of this dipole over geological timescales. The signatures of these polarity reversals are captured, for example, in magnetized rocks on the sea-floor. However, such reversal incidents have not been directly observed. In this subsection, we present the steady state configurations of the interactions between stellar wind and magnetospheres with no tilt and with extreme tilt angles ($45^{\circ}$ and $90^{\circ}$) and explain the reconnection schemes for both S-IMF and N-IMF cases. Thus, our simulations encompass the scenarios of: (a) planets with a highly tilted intrinsic magnetosphere, and (b) phases during field reversals modeled by a simple dipole whose inclination increases with time. We expore the tilt angles uptil $90^{\circ}$. This is because the relative orientation between stellar field and the dipolar axis, for inclinations ranging from $90^{\circ}-180^{\circ}$, are same as the orientations $0^{\circ}-90^{\circ}$ with the opposite $B_z$ in the IMF. Thus, we effectively capture a full reversal of the planetary dipole under the assumption that it remains dipolar during the reversal.

For each tilted dipolar magnetosphere, we incident the stellar wind and allow the simulations to reach a steady state. By steady state, we mean here the dynamical equilibrium configuration of the system at which the magnetic energy, kinetic energy and thermal energy become nearly invariant with time. Figure~\ref{fig:tilts_all_red} shows the steady state magnetospheric configurations for the above mentioned cases. The colormaps are identified by the magnitude of the $y$-component of current density, i.e. the component perpendicular to the slices shown. The reconnection regions show higher current density. The magnetic field streamlines are plotted in 3D. If the streamlines are plotted on a 2D slice, we are actually restricting the streamline integration to occur only within the slice of an actually 3D data (magnetic field) which may give rise to unwanted artefacts.

Figure~\ref{fig:tilts_all_red}(a) shows the steady state magnetosphere for no tilt and S-IMF case as a standard for comparison. The day-side dipolar lobe is heavily compressed with a reconnection zone at the substellar point due to which we observe high current density in this region. A mild intensity in current density, with respect to the background current, outlines the magnetospheric envelope due to the curvature of magnetic field in these regions. At the night-side, the pinching of planetary field lines leads to the formation of a magnetotail and purely stellar field lines at the right end of the box. The pinched region where reconnection occurs again shows higher current density. Thus, from our plots, we can associate the regions with high current density either with magnetic reconnection events or regions with highly curved magnetic field lines. In Fig.~\ref{fig:tilts_all_red}(b), the N-IMF case for no tilt magnetosphere is shown. This depicts a magnetosphere which is more complicated relative to the case of S-IMF. The highly compressed day-side lobe is indicated by the high current density. The reconnection zones, unlike the S-IMF case, is located on the northern and southern lobes on the night side magnetosphere. Mixed field lines either originate from the upper pole of the planet and leave through the lower boundary or they originate from the lower pole and leave through the upper boundary of the box. Purely stellar field lines enter through the lower boundary, curl around the extended magnetosphere and leave through the upper boundary.

In Fig.~\ref{fig:tilts_all_red}(c), a tilt of $45^{\circ}$ is introduced for the S-IMF case. As compared to the no tilt cases, this case shows a highly convoluted structure based on the trajectory of the magnetic field streamlines. The asymmetry in the day-side current density map is clearly visible in the form of a dark blue region bordering the tilted dipolar lobe. The sub-stellar reconnection zone is also shifted below the equatorial plane. The magnetotail is shifted above the equatorial plane along with the night-side reconnection zone. Purely stellar field lines form as a result of pinching of the planetary field lines and are found at the right end of the box (night-side). In the $45^{\circ}$ N-IMF case [Fig.~\ref{fig:tilts_all_red}(d)], very high current density is found in the north polar region. The magnetotail becomes very constricted due to the tilt. The reconnected field lines in this highly inclined dipole causes the open field lines to reach into the lower latitudes. This causes the reconnection events to induce plasma flow into lower latitudes.

In Fig.~\ref{fig:tilts_all_red}(e), the magnetic north pole faces an incoming S-IMF wind. The reconnections occur on the upper lobe leading to the formation of mixed field lines that leave the lower boundary on the day-side. Mixed lines originating from the magnetic south pole on the night-side leave through the right end of the box. A high current sheet region is found just below the equatorial plane where the pinching of planetary field lines takes place to give rise to purely stellar lines. The effect of an N-IMF wind on the $90^{\circ}$ magnetopshere would just be to flip the whole configuration about the horizontal. 

\section{Conclusions}\label{sec:con}
In this paper, we have created a star-planet interaction module using the 3D compressible MHD code PLUTO and studied the effects of the impact of stellar wind on a planetary magnetosphere for a ``far-out" star-planet system. The stellar wind is assumed to be a magnetized shock. We have considered the wind magnetic field to be either parallel (N-IMF) or anti-parallel (S-IMF) to the planetary dipolar axis. The magnetopause stand-off distance is matched with the expected value for various stellar wind velocities which serves the purpose of benchmarking our code.

A planetary magnetosphere with an Earth-like tilt is first considered to provide a point of comparison for increasingly complex configurations. The temporal evolution of the interaction reveals numerous features about the magnetic reconnections that take place on the route to a dynamical equilibrium. For S-IMF, the steady state magnetospheric configuration is not so complex. Reconnections occur at the day-side while a magnetotail is formed at the night-side due to pinching of planetary field lines and plasma blobs are sequentially ejected. On the other hand, for N-IMF, reconnections are mostly found at the polar regions and highly twisted field lines form at the night-side. Deviations from the original dipolar profile are found primarily at the night-side due to a large number of reconnection events in this region. The magnetic energy density with respect to that of the isolated dipolar magnetic field is found to be higher at the night-side due to clustering of field lines. The interesting features of the magnetotail dynamics demands a greater emphasis of spacecraft observations in this region to help us better understand the night-side interaction.

Injection of stellar and interplanetary plasma material into the planetary atmosphere may lead to the evolution of the planetary atmosphere with consequences for the habitability of the planet. We find the possibility of such inflow from the night-side magnetosphere. Our simulations show that inflow of stellar plasma into the inner magnetosphere is possible to within $x\approx 10R_{\rm E}$ through the magnetotail at the night-side. This plasma can then be trapped in the closed field lines of the planet and get sucked into the atmosphere as a consequence of the phenomenon of magnetic mirroring~\citep{kulsrud2005plasma}. It is to be noted here that in order to study magnetospheric injection, we have only considered large scale plasma flows and not particle dynamics (i.e., at the kinetic level). Such injection is known to lead to phenomenon such as auroras and radio emission -- which may be used to detect and characterize planetary or exoplanetary magnetospheres. We have also investigated atmospheric mass loss that occurs as a result of the interaction and have provided a quantitative estimate of mass loss rates in different directions. It is found that the loss is much higher in the night-side than the day-side as expected.

We have also considered magnetospheres with extreme tilt angles which is relevant for planets such as Neptune and Uranus in our solar system. This configuration can also mimic stages in planetary field reversals assuming that the profile remains dipolar during such polarity excursions. The interactions show that the magnetic reconnection regions are usually associated with large current density although high currents may also be found in regions of highly curved magnetic fields or clustered field lines due to wind impact. Global asymmetry in the magnetospheric structure is induced for such high tilts. The nature of reconnections are quite different for the S-IMF and N-IMF cases. The implications are low-latitude auroral formation and cosmic ray influx. Reconnection-powered radiation~\citep{2016ASSL..427..473U} serves as an important tool for identifying or detecting magnetospheric configurations in exoplanets. As deduced from our results, such radiation zones are expected to be found at higher or lower latitudes depending on the magnetospheric tilt. Therefore, confronting such simulations with observations -- whenever they become possible, would allow us to constrain the nature of exoplanetary magnetic dynamos and magnetospheres. 

The results of our numerical simulations pave the way for understanding how stellar winds shape planetary magnetospheres for different structural configurations and the role that magnetic reconnection plays in this process. We surmise that magnetohydrodynamic flow and magnetic reconnection mediated exchange of interplanetary plasma with that of planetary atmospheres would have implications for the evolution and habitability of planetary and exoplanetary systems.

\section*{Acknowledgement}
The development of the star planet interaction module was performed at the Center of Excellence in Space Sciences India and utilized computational resources funded by the Ministry of Human Resource Development, Government of India. SBD acknowledges the INSPIRE fellowship from the Department of Science and Technology, Government of India. BV would like to thank CESSI for the kind hospitality during his visit to IISER Kolkata as part of this work.

\bibliographystyle{aasjournal}

\end{document}